\documentstyle[twocolumn,aps]{revtex}

\begin{document}
\draft
\title{Shor-Preskill type security-proof for concatenated
Bennett-Brassard 1984 quantum key distribution protocol}

\author{Won-Young Hwang $^1$ \cite{email}
         Keiji Matsumoto $^1$ , Hiroshi Imai $^1$,
       Jaewan Kim $^2$, and Hai-Woong Lee $^3$} 

\address{$^1$ IMAI Quantum Computation and Information Project,
ERATO, Japan Science and Technology Corporation,
Daini Hongo White Bldg. 201, 5-28-3, Hongo, Bunkyo,
Tokyo 133-0033, Japan}

\address{$^2$ School of Computational Sciences,
Korea Institute for Advanced Study, Seoul 130-012, Korea }
 
\address{$^3$ Department of Physics,
Korea Advanced Institute of Science and Technology, Daejeon 
305-701, Korea }

\maketitle
\begin{abstract}
We discuss long code problem in the Bennett-Brassard
1984 (BB84) quantum key distribution protocol and describe
how they can be overcome by concatenation of the
protocol. 
Observing that concatenated modified Lo-Chau protocol
finally reduces to the concatenated BB84 protocol, we give the
unconditional security of the concatenated BB84 protocol.  
\end{abstract}
\narrowtext
\section{introduction}
Information processing with quantum systems enables
what seems to be impossible with its classical counterpart 
\cite{wies,shor,dowl,delg,hwan}.

Quantum key distribution (QKD) \cite{bene,eker,ben2,ben3}
is one of the most interesting and important
quantum information processings.  
QKD seems to become the first practical quantum information
processor \cite{gisi}.
 Although security of the Bennett-Brassard 1984 (BB84)
QKD
\cite{bene} had been widely conjectured based on the no-cloning 
theorem \cite{diek,woot},
it is quite recently that its unconditional security was
shown \cite{maye,biha,sho2}.
In particular, Shor and Preskill \cite{sho2}  showed the
security of BB84 scheme by elegantly using the connections
among several basic
ideas in quantum information processings, i.e. quantum error 
correcting codes (QECCs) \cite{cald,stea} and entanglement
purification \cite{ben4}.
\subsection{Long code problem and
concatenated BB84 protocol}
The Shor-Preskill proof affords security for a certain type of
 BB84 protocol 
 \cite{sho2} where we perform error correction and privacy
 amplification by a classical code associated with 
Calderbank-Shor-Steane (CSS) quantum code 
\cite{cald,stea,niel}.
 For a chosen code and a given error rate 
of the communication channels, Alice (the sender) and Bob
(the receiver) can calculate Eve's (Eavesdropper's) 
 information on the final key.
What Alice and Bob would want is to make Eve's information 
on the whole final key less than what they regard as negligible,
 for example, one bit. This can be done by choosing an 
appropriate code for the given error rate. 

However,
the Shor-Preskill proof does not place any bounds on the amount 
of effort Alice and Bob must do in decoding of the classical error 
correcting code associated with the CSS code  \cite{niel}.
In particular, when the code length is large the decoding is not so
 simple \cite{sho2}. One might say that we need not worry about 
the problem so much because we can do it by a code with
 a moderate length. However, we can easily see that it is risky:
 Let us consider case where the number of
 bits is $n$ and the error rate is $r$ ($0 \leq r \leq 1$). Then the
 probability distribution of the error rate will be peaked at $r$ with 
standard deviation $\sigma \sim \sqrt{r(1-r)/n}$. When the
 number of bits $n$ is one thousand
 and the error rate $r$ is $10\%$,
 the standard deviation is about $0.949\%$.
This means that the probability that the real error rate will be more
 than $12.4\%(=r+2.57 \sigma)$ is $1\%$. This implies 
that if they had assumed 
that a channel with $10\%$ error rate is
below a threshold, say $12.4\%$, for the one thousand bits, 
then the probability that Alice
and Bob will be cheated is $1\%$. Note that this is too
high a risk for a cryptographic task where we must achieve 
an exponentially
small probability to be cheated. Thus they have to
permit large room, say $20 \sigma$, between the error rate of
the channel and the threshold. Then the threshold must be larger
than $29.0 \%$ in this case.
However, no code is found to work yet if it
exceeds $26.4\%$  \cite{got2}.
Therefore, the number of bits $n$ must be
 large, say one million, in order to obtain the exponential 
security.
 However, as noted previously, decoding such a long code might
 be a difficult task \cite{sho2}.

On the otherhand, concatenation is a very useful and
 interesting idea in both classical and quantum error corrections 
 \cite{pres}. In this method, already-encoded-bits are used 
as unit-bits to encode new bits. The itrative processes
can be repeated as we like. The more they perform
the concatenation processes, the shorter encoded bits becomes
exponentially.
Since the concatenated codes are not included in the 
orginal codes, it is nontrivial one to adopt the concatenated
codes
for certain tasks. For example, it is by using concatenated 
coding that the fault tolerant quantum 
computation become possible \cite{got3}.  

In this paper, we show that the long code problem can be avoided
 by the idea of using concatenated BB84 protocol:
First, Alice and Bob generate raw keys. Next, using a classical
  codes associated with CSS code they perform 
 error correction and privacy amplification on the raw keys,
 as prescribed in the Ref. \cite{sho2}.
As a result they get first key.
The first key is not so much
 secure (possibly due to the long code problem). They
 perform the error correction and privacy amplification again in the 
same way on the first key and then they obtain  
the second key. They repeat this process until the estimated
leaked information on the final key is negligible. 

In section II, first we give the concatenated modified Lo-Chau
protocol. This protocol reduces to the
concatenated CSS code protocol that reduces to the
concatenated BB84 protocol.
In section III,
we discuss the security of the concatenated schemes and
how the concatenated protocol solves the long code problem.
In section IV, we give discussion and conclusion.
\subsection{Notation}
In this paper,  we use mostly the notations in Refs. \cite{sho2,lo3}.

The canonical basis of a qubit consists of 
$|0\rangle$ and $|1\rangle$. We define another basis as follows.
$|\bar{0}\rangle= (1/\sqrt{2})(|0\rangle+|1\rangle)$ and
$|\bar{1}\rangle= (1/\sqrt{2})(|0\rangle-|1\rangle)$.
$H$ is the Hadamard transform.
This transformation interchanges the bases
$|0\rangle$, $|1\rangle$ and $|\bar{0}\rangle$, $|\bar{1}\rangle$. 
 $I= \sigma_0$ is the identity operator and
$\sigma_x$, $\sigma_y$, $\sigma_z$
are the Pauli operators.
The $\sigma_{{a}(i)}$ denotes the Pauli operator $\sigma_{a}$ 
acting on the $i$-th qubit where $a=0,x,y,z$.
For a binary vector $s$, we let
$\sigma_a^{[r]}= 
\sigma_{{a}(1)}^{s_1} \sigma_{{a}(2)}^{s_2}
 \cdot \cdot \cdot \sigma_{{a}(n)}^{s_n} $,
 where $s_i$ is the $i$-th bit of $s$ and
$\sigma_a^0= I$, $\sigma_a^1=\sigma_a $.

The Bell basis are the four maximally entangled state,
$|\Psi^{\pm}\rangle =(1/\sqrt{2})(|01\rangle \pm |10\rangle)$
and
$|\Phi^{\pm}\rangle =(1/\sqrt{2})(|00\rangle \pm |11\rangle)$.
 
Let us consider two classical binary codes, $C_1$ and $C_2$,
such that $\{0\} \subset C_2 \subset C_1 \subset F_2^n $
 where
$ F_2^n $ is the binary vector space of the $n$ bits.
A set of basis for the CSS code can be obtained from vectors
$v \in C_1$ as follows,
$v \rightarrow (1/|C_2|^{1/2}) \sum_{w \in C_2}|v+w \rangle $.
Note that $v_1$ and $v_2$ give the same vector if
$v_1-v_2 \in C_2$.
$H_1$ is the parity check matrix for the code $C_1$ and
$H_2$ is that for $C_2^\perp$, the dual of $C_2$.
 $Q_{x,z}$ is a class of QECCs. For $v \in C_1$, the corresponding
 code word is
 $v \rightarrow (1/|C_2|^{1/2}) \sum_{w \in C_2}
  (-1)^{z \cdot w}|x+v+w \rangle$.
\section{concatenated BB84 protocol}
First,  we give the concatenated modified Lo-Chau protocol.
This scheme reduces to the concatenated CSS codes protocol
that reduces to the concatenated BB84 protocol.

We consider a doubly concatenated scheme of 
the $[[n_1,k_1,d_1]]$ and  $[[n_2,k_2,d_2]]$  CSS codes
\cite{cald,stea}. It is clear that it can be generalized to
multiple-concatenation in the same ways.

{\it Protocol A:
Concatenated modified Lo-Chau protocol.}

{\it The first stage}:
(1) Alice creates $2 n_1 n_2$ EPR pairs in the state
$|\Phi^+\rangle ^{\otimes (2 n_1 n_2)}$. 
(2) Alice selects a random  $(2 n_1 n_2)$-bit string $b$.
 She performs a Hadamard operation on second half of each 
EPR pair for which the component of  $b$ is one.
(3) Alice sends the second half of each EPR pair to Bob.
(4) Bob receives the qubits and publically announces this fact.
(5) Alice announces the bit string $b$.
(6) Bob undoes the Hadamard operations in step 2.
(7) Alice selects randomly $ n_1 n_2$ of the 
$2 n_1 n_2$ EPR pairs to serve as check-bits.
(8) Alice and Bob each measure their halves of the 
$ n_1 n_2$ check EPR
pairs in the $ \{ |0\rangle,|1\rangle \} $  basis
and share the results. If too many of these measurements
disagree, they abort the scheme
(9)  Alice randomly chooses $n_1$ unencoded qubits. 
She repeats this until all code bits are used. As a result,
she gets $ n_2$ sets of code bits
 whose number of elements is all $n_1$. 
 She announces her choices to Bob.
(10) For each set, they make the measurements of
$\sigma_z^{[r]}$ for each row $r \in H_1$ and  
$\sigma_x^{[r]}$ for each row $r \in H_2$ of the $[[n_1,k_1,d_1]]$
CSS code. 
Alice and Bob share the results, compute the syndromes for bit
and phase flips, and then
transforms their state so as to obtain $ k_1 n_2 $ 
once-encoded high-fidelity EPR pairs. 

In the next stage, Alice and Bob do essentially the 
same operations on 
the $ k_1 n_2 $  once-encoded EPR pairs as they have done
on un-encoded EPR pairs in the previous stage.

 {\it The second stage}:
(1) Alice randomly chooses $n_2$ once-encoded qubits. 
She repeats this until
all code bits are used. As a result, there are $ k_1$ sets whose
number of elements is $n_2$. She announces her choices to Bob.
(2) For each set, Alice and Bob make
the measurements of $\sigma_z^{[r]}$ for each row $r \in H_1$ 
and  
$\sigma_x^{[r]}$ for each row $r \in H_2$ of the $[[n_2,k_2,d_2]]$
CSS code. They share the
results, compute the syndromes for bit and phase flips, and then
transforms their state so as to obtain $k_1 k_2$ 
doubly-encoded EPR pairs.
(3) Alice and Bob measure the EPR pairs in the
doubly encoded $\{|0\rangle, |1\rangle\}$
basis to obtain $k_1 k_2$-bit final key.
$\Box$

Let us now consider reduction of the protocol. The same
arguments used in Ref. \cite{sho2} applies here.
The difference is that they measure the syndrome twice at
both level of concatenation, that is, un-encoded qubits and 
once-encoded qubits.  

{\it Protocol B: Concatenated CSS code protocol.}

{\it The first stage}: 
(1) Alice creates $ n_1 n_2 $ random check bits
and a random $(2 n_1 n_2)$-bit string $b$.
(2) Alice chooses $ n_2$  $n_1$-bit string $x_i$
 and $z_i$ at random ($i= 1,2,..., n_2$).
(3) Alice prepares $k_1 n_2 $ once-encoded bits in a state 
$|00\cdot \cdot \cdot 0\rangle$ using $[[n_1,k_1,d_1]]$ CSS 
code $Q_{x_i,z_i}$.

{\it The second stage}:
(1) Alice creates a random
 $k_1 k_2$-bit key-string $k^{\prime}$.
(2) Alice chooses $k_1$  $n_2$-bit string 
$x^{\prime}_j$ and $z^{\prime}_j$ at random ($j= 1,2,...,k_1$).
(3) Alice encodes  her key  $|k^{\prime}\rangle$
using the once-encoded qubits prepared in the step 3 of the 
first stage with the  $[[n_2,k_2,d_2]]$ CSS code 
$Q_{x^{\prime}_j, z^{\prime}_j}$.
(4) Alice randomly chooses $ n_1 n_2$ positions out of 
$2 n_1 n_2$ and puts the unencoded check bits in these
 positions. 
She randomly permutes the qubits of the doubly encoded code
bits of the previous step. She puts them in the remaining positions.
(5) Alice performs the Hadamard operation on each
qubit for which the component of  $b$ is one.
(6) Alice sends the resulting state to Bob. Bob acknowledge
the receipt of the qubits.
(7) Alice announces the string $b$, the positions and values 
of unencoded check bits, the random permutation
and each $x_i$, $z_i$ and $x^{\prime}_j$, $z^{\prime}_j$.
(8) Bob undoes the Hadamard operation and random permutation.
(9) Bob measures unencoded check-bits
in the $ \{ |0\rangle,|1\rangle \} $  basis
and announces the results to Alice.
If too many of these measurements disagree,
they abort the scheme.
(10) Bob measures the qubits in the 
doubly encoded $\{|0\rangle, |1\rangle\}$ basis to obtain 
$k_1 k_2$-bit final key with near-perfect security. $\Box$

Let us consider reduction of
the concatenated CSS code protocol to the concatenated
BB84 protocol. 
We can apply the same arguments used in Ref. \cite{sho2}.
The difference is that  reductions 
process involved with Eq. (4) in Ref. \cite{sho2} are performed
twice at both level of concatenation.

{\it Protocol C: Concatenated BB84 protocol.}

{\it The first stage}:
(1) Alice creates $4 n_1 n_2(1+\delta)$ random bits.
(2) Alice chooses a random $4 n_1 n_2(1+\delta)$-bit
 string $b$. For each bit, she creates a state in the 
$ \{ |0\rangle,|1\rangle \} $ and 
$\{ |\bar{0}\rangle, |\bar{1}\rangle \}$ basis,
if the corresponding component of the bit
$b$ is zero and one, respectively.
(3) Alice sends the resulting qubits to Bob.
(4) Bob receives the $4 n_1 n_2(1+\delta) $ qubits, 
measuring each in the  $ \{ |0\rangle,|1\rangle \} $ and 
$\{ |\bar{0}\rangle, |\bar{1}\rangle \}$ basis 
at random.
(5) Alice announces $b$.
(6) Bob discards any results where he measured a different basis
than Alice prepared. With high probability, there are at least
$2 n_1 n_2 $ bits left. Alice decides randomly on a set of 
$2 n_1 n_2 $ bits to use for the protocol, and chooses at
 random $ n_1 n_2$ of these to be check-bits.
(7) Alice and Bob announce the values of the their check-bits. If 
too few of these values agree, they abort the protocol.
(8) Alice announces $u+v$, where $v$ is a string consisting of
randomly chosen code-bits, 
and $u$ is a random code word in $C_1$ of the $[[n_1,k_1,d_1]]$ 
code. Alice announces the $n_1$ positions of the randomly
 chosen  code-bits.
(9) Repeat the previous step $ n_2$ times,
until all code-bits are consumed.
(10) Bob subtracts each
$u+v$ from each of his code-bits, $v+\epsilon$, and
corrects the result, $u+\epsilon$, to a codeword in $C_1$ of 
the $[[n_1,k_1,d_1]]$ code.
(11) Alice and Bob use the coset of each $u+C_2$ as the key.
In this way, they obtain $ k_1 n_2$-bit string.

{\it The second stage}:
(1) Alice announces $u+v$, where $v$ is a string consisting of
randomly chosen code-bits, 
and $u$ is a random code word in $C_2$ of the $[[n_2,k_2,d_2]]$ 
code. Alice announces the $n_2$ positions of the randomly
chosen code-bits.
(2) Repeat the previous step at $ k_1$ times,
until all code-bits are consumed.
(3) Bob subtracts each 
$u+v$ from his each code qubits, $v+\epsilon$, and
corrects the result, $u+\epsilon$, to a codeword in $C_1$ of the 
$[[n_2,k_2,d_2]]$ code.
(4) Alice and Bob use the coset of each $u+C_2$ as the key.
In this way, they obtain $k_1 k_2$-bit key. $\Box$ 
\section{security of the protocol}
What we want to show is the security of the protocol C, the
cancatenated BB84 protocol. However,
it is sufficient for us to show the security of the protocol $A$,
since the protocol $A$ reduces to the protocol $C$.
Accordingly arguments in the following are for the protocol $A$ 
that accompanies entanglement purification.

The security of modified Lo-Chau protocol is based on the
idea of random sampling \cite{sho2,lo2}:
Based on the measured error rate in the check-bits, Alice 
and Bob estimate actual error rate in the code-bits. 
Since the check-bits are 
randomly chosen, they can do it with a high reliability. 
Thus they can correct errors with a high reliability.
(It is not misleading to use notations of QECCs in the discussion
of entanglement purification
because they are equivalent to each other here.)

It is not difficult to intuitively understand how the protocol 
$A$ is secure. When the estimated actual error rate in the 
code-bits is not low enough, they cannot obtain EPR pairs with
a fidelity that is satisfactorily high. However, in most cases the
fidelity has become higher.
By iterating the entanglement purification, 
they can make the fidelity become
higher and higher. 
The protocol $A$ is secure when EPR pairs have high enough fidelity,
since the almost perfect fidelity implies the
almost perfect security \cite{lo,sho2}.
This is indeed the original idea of quantum privacy amplification
\cite{deut}.

Let us give a more detailed description.
The probability distribution of the number of actual errors forms
an approximate Gaussian distribution whose deviation depends 
on the length of code.
The longer the codes is, the smaller the deviation is. 
A rigorous relavant equation regarding random sampling 
for this estimation is given in Ref. \cite{lo2}.
Let us consider the example considered in introduction.
The error rate $r$ is $10\%$. The number of code-bits $n$ is
one thousand and thus the standard deviation is about $0.949\%$.
This means that the probability that the real error rate will be less
than $12.4\%(=r+2.57 \sigma)$ is $99\%$.
Therefore, if the threshold of the error correcting code 
is greater than $12.4\%$, the error rate of purified EPR pairs will 
be less than $1\%$, that is much smaller than the original error
rate $10\%$.
By repeating this operation, the fidelity can be made arbitrily small.
Let us estimate how error rate $r$ decreases with iteration of
purification. For a given length of code $n$, the deviation $\sigma$
is approximately proportional to $\sqrt{r}$. (See introduction.)
Let us denote $i$-th error rate and deviation by $r_i$ and
$\sigma_i$, respectively, where $i$ is positive integer.
Then we can get a relation 
$r_{i+1}\sim \exp(-T^2/\sigma_i^2) = \exp(-T^2/r_i)$ 
when $r\sim 0$. Here constant $T$ is the threshold of the code.
We can see that $i$-th error rate $r_i$ super-exponentially
decreases  with respect to $i$. Fidelity is proportional to
$1-r$ and thus almost perfect fidelity can be obtained.

In the protocol $A$, Alice randomly chooses sets 
of qubits to be used for syndrome measurement. This corresponds
to random choice of code-bits to be used for error correction and
privacy amplification (in the step 8 of the first stage
and the step 1 of the second stage) in the Protocol $C$.
It should be noted that the randomness in the choice is 
essential in the proposed protocol: The error rate is what is
averaged for all $n_1 n_2$ check-bits. Unless the choice
for the code-bits is random, Eve can succesfully
cheat by correlating positions of the errors in code-bits.
\section{conclusion} 
In a certain class of  BB 84 protocol with proven security
\cite{sho2},  long classical error correcting
codes are desired in order to obtain 
sufficient security. However, it is not easy to decode long codes.
If they use intermediate length codes they cannot
obtain enough security. We have shown that this long code 
problem can be resolved by concatenation of the BB 84 protocol.
We have shown the security of the concatenated BB84 protocol:
The concatenated BB84 protocol can be
derived from the concatenated modified Lo-Chau protocol.  
We have described how the latter protocol is secure.
\acknowledgments
W.-Y. H., M. K., and H. I. are very grateful to 
Japan Science Technology Corporation for financial supports
and to Dr. Wang Xiang-Bin for helpful discussions.
H.-W. L. appreciates the financial support from the Brain Korea 21
Project of Korean Ministry of Education. J.K. was supported by
Korea Research Foundation Grant 070-C00029. 


\begin{references}
\bibitem[*]{email}  Present address: 
Center for Photonic Communication and Computing,
Northwestern University,
2145 N Sheridan Road, Evanston, IL 60208, USA;
Email address: wyhwang@ece.northwestern.edu
\bibitem{wies} S. Wiesner, Sigact News {\bf15}(1), 78  (1983).
\bibitem{shor} P. Shor, Proc. 35th Ann. Symp. on Found. of 
               Computer Science. (IEEE Comp. Soc. Press,
               Los Alomitos, CA, 1994) 124-134.
\bibitem{dowl} J.P. Dowling, Phys. Rev. A {\bf 57}, 4736 (1998).
\bibitem{delg} A. Delgado, W.P. Schleich, and G. S\"{u}ssmann,
                New Journal of Physics {\bf 4}, 37.1 (2002).
\bibitem{hwan} W.Y. Hwang, D. Ahn, S.W. Hwang, and Y.D. Han,
                Eur. Phys. J. D {\bf 19}, 129 (2002).
\bibitem{bene} C.H. Bennett and G. Brassard, in : Proc. IEEE
                Int. Conf. on Computers, systems, and signal
                processing, Bangalore (IEEE, New York, 1984)
                p.175.
\bibitem{eker} A.K. Ekert, Phys. Rev. Lett. {\bf67}, 661 (1991).
\bibitem{ben2} C.H. Bennett, G. Brassard, and N.D. Mermin, Phys.
               Rev.Lett. {\bf68},  557 (1992).
\bibitem{ben3} C.H. Bennett,
               Phys. Rev. Lett. {\bf68}, 3121  (1992).
\bibitem{gisi} N. Gisin, G. Ribordy, W. Tittel, H. Zbinden,
               Rev. Mod. Phys. {\bf 74}, 145 (2002), 
               references therein.
\bibitem{diek} D. Dieks, Phys. Lett. A {\bf92}, 271 (1982).
\bibitem{woot} W.K. Wootters and W.H. Zurek, Nature
                {\bf299}, 802  (1982).
\bibitem{maye} D. Mayers,
               J. Assoc. Comput. Mach. {\bf 48}, 351 (2001).
\bibitem{biha} E. Biham, M. Boyer, P.O. Boykin, T. Mor, and
               V. Roychowdhury, in {\it Proceedings of the
               Thirty-Second Annual ACM Symposium on Theory
               of Computing} (ACM Press, New York, 2000),
               pp.715-724, quant-ph/9912053.
\bibitem{sho2} P.W. Shor and J. Preskill, Phys. Rev. Lett.
               {\bf 85}, 441 (2000).
\bibitem{cald} A.R. Calderbank and P.W. Shor, Phys. Rev. A
               {\bf54}, 1098  (1996).
\bibitem{stea} A.M. Steane, Phys. Rev. Lett. {\bf77},
               793 (1996).
\bibitem{ben4} C.H. Bennett, D.P. DiVincenzo, J.A. Smolin, and
               W.K. Wootters, Phys. Rev. A {\bf54},
               3824  (1996).
\bibitem{niel} M.A. Nielsen and I.L. Chuang, {\it Quantum   
               Computation and Quantum Information} (Cambridge 
               University Press, Cambridge, 2000).
\bibitem{got2} D. Gottesman and H.-K. Lo, quant-ph/0105121.
\bibitem{pres} J. Preskill, quant-ph/9705031.
\bibitem{got3} D. Gottesman, quant-ph/9903099.
\bibitem{lo3} H.-K. Lo, Quan. Inf. Com. {\bf 1}, 81 (2001).
\bibitem{lo2} H.-K. Lo, H.F. Chau, and M. Ardehali, 
               quant-ph/0011056.
\bibitem{lo}  H.-K. Lo and H.F. Chau, Science {\bf283},
             2050 (1999).
\bibitem{deut} D. Deutsch, A. Ekert, R. Jozsa, C. Macchiavello,
               S. Popescu, and A. Sanpera, Phys. Rev. Lett.
               {\bf 77}, 2818 (1996): {\bf 80}, 2022 (E) (1996).

\end{references}
\end{document}